\documentclass[twocolumn]{aastex63} 

\usepackage{amsmath}

\revised{\today}
\submitjournal{ApJ}

\shorttitle{Exoclimate Model Errors}
\shortauthors{Kopparla et al.}
\begin{document}

\title{General Circulation Model Errors are Variable across Exoclimate Parameter Spaces}

\correspondingauthor{Pushkar Kopparla}
\email{pushkarkopparla@gmail.com}

\author[0000-0002-8951-3907]{Pushkar Kopparla}
\altaffiliation{CSH Fellow}
\affiliation{Center for Space and Habitability, 
University of Bern, Switzerland}

\author[0000-0001-9423-8121]{Russell Deitrick}
\affiliation{Center for Space and Habitability, 
University of Bern, Switzerland}

\author[0000-0003-1907-5910]{Kevin Heng}
\affiliation{Center for Space and Habitability, 
University of Bern, Switzerland}
\affiliation{Department of Physics, University of Warwick, U.K.}
\affiliation{Ludwig Maximilian University, University Observatory Munich, Scheinerstrasse 1, Munich D-81679, Germany}

\author[0000-0002-6907-4476]{Jo\~{a}o M. Mendon\c{c}a}
\affiliation{National Space Institute, Technical University of Denmark, Denmark}

\author[0000-0002-6893-522X]{Mark Hammond}
\affiliation{Department of Geosciences, 
University of Chicago,  USA}



\begin{abstract}

General circulation models are often used to explore exoclimate parameter spaces and classify atmospheric circulation regimes. Models are tuned to give reasonable climate states for standard test cases, such as the Held-Suarez test, and then used to simulate diverse exoclimates by varying input parameters such as rotation rates, instellation,  atmospheric optical properties, frictional timescales and so on. In such studies, there is an implicit assumption that the model which works reasonably well for the standard test case will be credible at all points in an arbitrarily wide parameter space. Here, we test this assumption using the open-source general circulation model THOR to simulate atmospheric circulation on tidally locked Earth-like planets with rotation periods of 0.1 to 100 days. We find that the model error, as quantified by the ratio between physical and spurious numerical contributions to the angular momentum balance, is extremely variable across this range of rotation periods with some cases where numerical errors are the dominant component. Increasing model grid resolution does improve errors but using a higher-order numerical diffusion scheme can sometimes magnify errors for finite-volume dynamical solvers. We further show that to minimize error and make the angular momentum balance more physical within our model, the surface friction timescale must be smaller than the rotational timescale. 

\end{abstract}



\section{Introduction} \label{sec:intro}

General circulation models (GCMs) are a powerful tool to explore the space of possible climates on exoplanets, especially to capture three-dimensional, multi-component and non-linear interactions within the climate system which are hard to study with simpler, low-dimensional models \citep{kaspi2015atmospheric}. One of the main takeaways from exoclimate parameter space studies has been that transitions between different flow regimes can be understood by examining trends in circulation with parameters such as planetary rotation rate, strength of irradiation or drag \citep[for eg.,][]{mitchell2010transition,pascale2013nonequilibrium,pierrehumbert2019atmospheric}. 

Modern GCMs can produce atmospheric circulations that are qualitatively similar to observations for solar system planets \citep{read2016global}, but there remain many unsolved challenges particularly in predicting time variability, the most famous instance being the decades long effort to accurately model climate change on Earth \citep{eyring2016overview}. GCMs are imperfect models limited by grid resolution and finite timesteps, employing parameterizations for complex processes which are either too computationally expensive to represent accurately or not well understood physically. Therefore, there are significant omissions, biases and uncertainties involved \citep{heng2011atmospheric,amundsen2014accuracy}. In the presence of high volumes of observational data, such as in the case of the Earth, GCMs can be improved iteratively to yield the right spatial and temporal variability \citep{koutroulis2016evaluation}. For exoplanets, time-variability is now beginning to be observed, such as the hot Jupiters Kepler-76b \citep{jackson2019variability}, WASP-12b \citep{hooton2019storms} and the super Earth Cancri 55e \citep{tamburo2018confirming}.  The spatial and temporal patterns associated with this variability could help us calibrate our exoclimate GCMs, but sufficiently high-resolution observations may never be possible for many exoplanets, given the large uncertainties in physical properties \citep{heng2015atmospheric}. Thus, it is not clear how well our GCMs can capture the processes that drive exoclimate variability.

When direct model validation is not possible due to data scarcity, we rely on inter-model comparisons to understand if the model results are credible, whether it be between a GCM and a hierarchy of simpler models \citep{liu2013atmospheric,zhang2017effects} or between multiple GCMs using standard test setups \citep{heng2011atmospheric,polichtchouk2014intercomparison}, such as the well known Held-Suarez test \citep{held1994proposal} used commonly for Earth system models. In case of disagreements between two GCMs, Held-Suarez type standard tests cannot judge which GCM is more correct because they only check for qualitative characteristics of the flow, so other metrics become necessary to determine model credibility. 

There is a precedent for such a metric in planetary science: a study using identical forcings in different GCMs gave differing results for simulations of Venus' superrotating atmosphere, indicating dependencies on the nature of the dynamical core used and in particular the parameterization of the numerical diffusion \citep{lee2010general}. Further investigations showed that atmospheric angular momentum conservation varied significantly between models \citep[henceforth Leb12]{lebonnois2012angular} which explained why there were differences in circulation and the models with the best angular momentum conservation properties produced the most credible atmospheric states \citep{lee2012angular}. Since tidally-locked terrestrial exoplanets are expected to have atmospheres dominated by a superrotating jet \citep{hammond2020equatorial}, arguments relating angular momentum conservation to realism of the model circulation for Venus can plausibly be extended to these exoplanets as well.

In this paper, we explore model errors in the simulation of Earth-like tidally locked planets with different rotation rates. The paper is organised as follows: Section \ref{sec:methods} describes the model setup and angular momentum budget calculations.  Section \ref{sec:methods} after that outlines the trends in angular momentum variability and their origins.  Section \ref{sec:results} summarizes the findings and suggests directions for future research.

\section{Methods and Model Setup }
\label{sec:methods}
In this work we use THOR, an open source general circulation model, for our simulations \citep{mendonca2016thor,deitrick2020thor}. It solves the non-hydrostatic Euler equations on a sphere with a finite-volume solver \citep{tomita2001shallow} using an icosahedral grid. The simulation setup is the tidally-locked Earth benchmark case described in Sec 4.1 of \citet{deitrick2020thor}, which was first proposed in \citet{heng2011atmospheric} and is based on the Held-Suarez test commonly used for Earth GCM studies \citep{held1994proposal} with a tidally-locked forcing function. The planet has a radius equal to the Earth, with a one bar atmosphere and a mean molecular mass of 29 $g/mol$. There are no hazes or clouds, the forcing is a Newtonian cooling scheme with convective adjustments enabled and surface friction is implemented using a Rayleigh friction scheme with a frictional timescale of one day. The grid resolution is about $4^\circ$ \cite[see Eqn. 3 of ][]{mendonca2016thor}. The model time step is 600$s$ and each \replaced{model run}{simulation} is for 2400 days with outputs being \replaced{written out}{sampled} every 24 days. The numerical diffusion flux is calculated using a fourth-order hyperdiffusion scheme, called so because it contains two Laplacians. The general form is:

\begin{equation}
    F_{\phi} = (-1)^{n/2 + 1} \nabla^2_h \left[ \mu K_h (\nabla_h^{n-2} \phi ) \right],
\end{equation}

where $n$ is the order of the scheme and has a value of 4, and 

\begin{eqnarray*}
     (\mu, \phi) = \begin{cases}
                            (1, \rho)  & \text{Continuity equation},\\
                            (\rho, \vec{v}_h) & \text{Horizontal momentum equations}, \\
                            (\rho, v_r) & \text{Vertical momentum equation}, \\
                             (R_d \rho, T) & \text{Thermodynamic equation}. \\
                            \end{cases} 
\end{eqnarray*}

where $\rho$ is the density, $\vec{v} _h$ are the horizontal velocities, $v _r$ is the radial (vertical) velocity, $R_d$ is the specific gas constant and $T$ is the temperature. $K_{h}$ is the diffusive coefficient which scales with grid spacing $d$ and model timestep $\Delta t$ as \citep{deitrick2020thor}:

\begin{equation}
    K_{h} = D_{h}\frac{d^4}{\Delta t}
\end{equation}

$D_{h}$ is a non-dimensional diffusion constant with a baseline value of 0.0048. It is one of the main tuning parameters responsible for model stability.

Since the model equations explicitly conserve linear momentum but not angular momentum, the model angular momentum conservation properties can then be used as a measure of model correctness, as argued by Leb12. Following their convention, the total atmospheric angular momentum budget can be decomposed as follows:

\begin{eqnarray}
    M &=& M_o+M_r \nonumber \\
      &=& \int_V \omega a^2 \cos^2\theta\rho dV + \int_V u\,a\cos\theta \rho dV
\end{eqnarray}

The first term, $M_o$, is an integral over the entire atmosphere $V$ capturing the effects of solid body rotation, where $\omega$ is the rotation rate of the planet, $a$ is the planet radius, $\theta$ is the latitude,  $\rho$ is the density, $g$ is the acceleration due to gravity. The second term, $M_r$, is the relative angular momentum resulting from zonal wind speeds $u$ measured relative to the surface, integrated over the volume of the atmosphere $V$. We have recast the expression for $M_o$ which used a surface pressure term in Leb12 into a non-hydrostatic form more suitable for the present model. The contributions to this angular momentum within the model can be broken down as below:

\begin{equation}
    \frac{dM}{dt} = \frac{dM_o}{dt}+\frac{dM_r}{dt} = F + T + D + S + \epsilon
\end{equation}

The $\frac{dM_o}{dt}$ term accounts for contributions from surface pressure changes. The $\frac{dM_r}{dt}$ term is from changes in the zonal wind speed, $F$ is the surface friction contribution, $T$ is the topographic or mountain torque, $D$ is the torque due to drag from numerical diffusion, $S$ is from the sponge layer and $\epsilon$ are the residual conservation errors in the dynamical core. The above equation is written for global net torques which are summed over the entire volume of the atmosphere. Thus, if $F_{loc}$,  $\epsilon _{loc}$ and $D_{loc}$ are the physical, dynamical error and numerical diffusion related torques evaluated at each grid point, we can write:

\begin{eqnarray}
    F &=& \Sigma F_{loc} = \Sigma  \left( \frac{du}{dt} \right)_F  a \rho dV \nonumber \\
\end{eqnarray}

where the summation $\Sigma$ is over all grid points in the model domain,  $\left( \frac{du}{dt} \right)_F$ is the change in zonal velocity at the grid point due to frictional forces and $dV$ is the volume of the grid cell. The expressions for the summations of $D_{loc}$ and $\epsilon_{loc}$ follow the same pattern.

 Within our model setup, there is no topography or sponge layer \citep{jablonowski2011pros}, and we can ignore the small $\frac{dM_o}{dt}$ term (which is about 2-3 orders of magnitude smaller than $\frac{dM_r}{dt}$) to write a simplified expression for the numerical errors in the dynamical core:

\begin{equation}
    \epsilon = \frac{dM_r}{dt} - F - D 
    \label{eqn:DynError}
\end{equation}

A new functionality was added to THOR to output wind speeds at different points during each model timestep - at the start of the timestep, after the physics module and at the end of the timestep (i.e., after the dynamical core step). The difference in wind speeds between the start and end of the timestep is used to calculate $\frac{dM_r}{dt}$, while the change in wind speed between the start and after the physics module gives the surface friction contribution $F$. The numerical diffusion contribution to the wind speed ($\delta u_D$) was also separately written to output, yielding $D$.  Then the model error $\epsilon$ is calculated from Eqn. \ref{eqn:DynError}. Thus, we have:

\begin{eqnarray}
\left( \frac{du}{dt} \right)_F &=& \frac{u_{mid}-u_{start}}{\delta t} \nonumber \\
\left( \frac{du}{dt} \right)_\epsilon &=& \frac{u_{end}-u_{mid}-\delta u_{D}}{\delta t}
\end{eqnarray}

where $\delta t$ is the timestep, $u_{start}$ is the zonal wind at the start of the timestep, $u_{mid}$ is after the physics module (friction) is calculated and $u_{end}$ is after dynamics calculations and is at the end of the timestep.

In a perfect model, eventually the atmospheric angular momentum equilibrates and the net torque should become vanishingly small, but this rarely happens in practice due to finite simulation times and long period atmospheric and/or numerical oscillations. The ratio between the absolute value of the physical ($F$) and spurious numerical ($\epsilon$+D) torques is calculated at each output timestep over the run-time. The time median of this ratio over the entire simulation time gives the Global Credibility Score, $GCS$: 

\begin{equation} \label{eqn:gcs}
    GCS =  median\left({\left| \frac{F}{\epsilon+D} \right|}\right) = median\left( \left| \frac{\Sigma F_{loc}}{ \Sigma(\epsilon_{loc} +D_{loc})}\right|\right)
\end{equation}

Simulations where $\epsilon + D$ is equal to or larger than $F$ on average or equivalently $GCS \leq 1$ are totally not trustworthy, and $GCS>10$ is preferable so that physical torques are at least an order of magnitude larger than numerical error.  See Appendix A for more details on this score. Taking ratios at individual grid points can therefore blow up to arbitrarily large values, which is why we first sum over all grid points before calculating the ratio. 

However, this global average metric does not give the full picture. If for instance, there are large surface frictional torques of very similar magnitude but eastward at the tropics and westward in the midlatitudes, the global surface torque sum would be very small even though these torques may significantly affect local flows. Thus, the above GCS metric might indicate that the model is unphysical, even if the local physical torques were quite large compared to numerical errors.  We recognize that this is a shortcoming of this metric.
\section{Results and Discussion}
 \label{sec:results}
\begin{figure*}
 \plotone{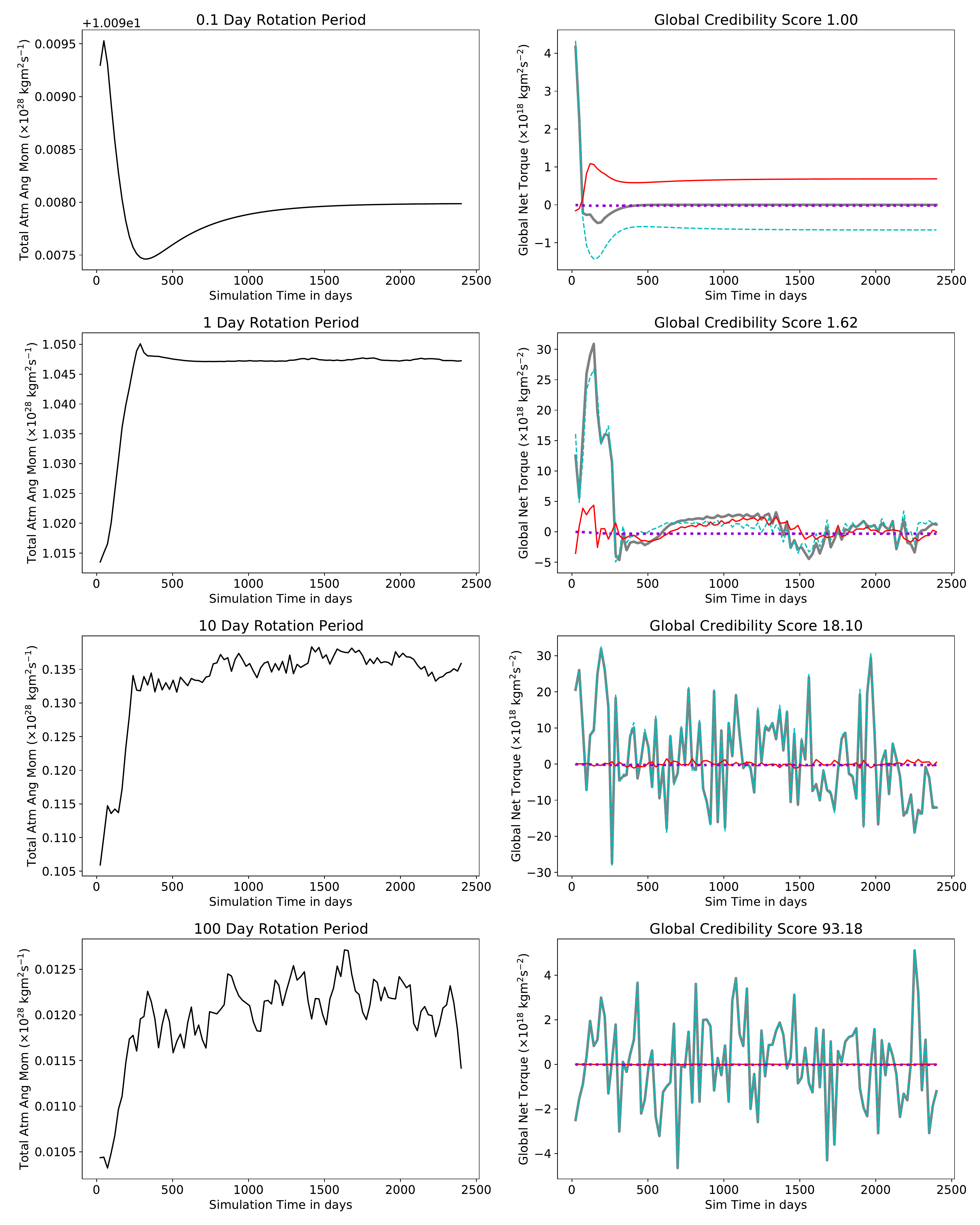}
 \caption{Evolution of the total atmospheric angular momentum (left panel) and total torque ($dM_r/dt$) and its contributions from surface friction, numerical dissipation and conservation errors in the dynamical core for planets with rotation rates from 0.1 to 100 days. The quantities plotted in the right column are global net torques. Simulations with higher scores are more credible.\label{fig:MainErrors}}
 \end{figure*}

Fig. \ref{fig:MainErrors} shows the results of the above described angular momentum analysis with the GCS metric for four planets with different rotation rates. After a spin-up period of a few hundred days, the angular momentum reaches a steady state with some variability about a mean value. For slow rotators, angular momentum changes correspond to changes in physical torques like surface friction contributions and dynamical error is negligible. But as rotation rate increases, conservation errors from the dynamical core become larger and eventually for the fastest rotators, the equilibrium is set by a balance between the physics and the dynamical errors of nearly equal magnitude. So the model goes from being quite credible to very unreliable over this range.

In the absence of large physical torques such as mountain torque, numerical torques being prominent is somewhat expected \citep{lebonnois2012angular,lauritzen2014held}. Since exoplanet simulations rarely have topography because we have never observed an exoplanet surface, this problem is likely to persist. The global averages seem to reduce numerical torques more than physical ones, leading to a larger range of values for the GCS.

\begin{figure*}
 \plotone{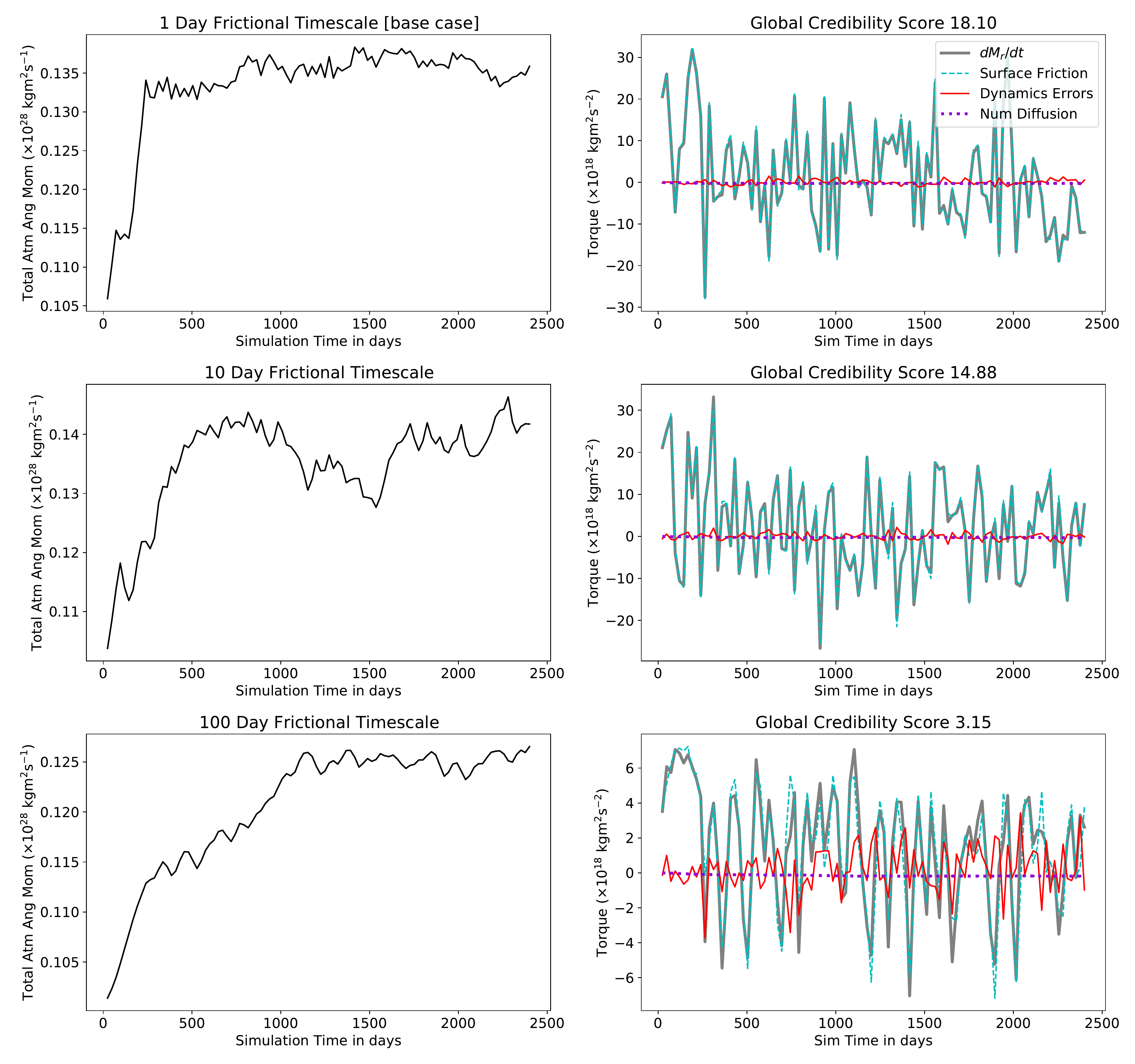}
 \caption{Same as Fig. \ref{fig:MainErrors} but with different frictional timescales for a rotation rate of 10 days.} \label{fig:FricErrors}
 \end{figure*}

 Since there is no sponge layer in this model run, the physical sources and sinks for atmospheric angular momentum are at the surface. For the slow rotators, the frictional time scale acts to dissipate energy on short time scales (i.e., shorter than the rotational period).  This energy is then lost, and it is artificially replenished by the forcing. For the fast rotators, when surface friction acts too slowly to adjust the atmosphere to the surface leading to low variability, the dynamical errors take over the torque balance. Since the frictional timescale is an important control on this process, we varied it \replaced{to}{between} 10 and 100 days from the baseline value of 1 day to test the model error sensitivity. As the frictional timescale becomes slower, the credibility scores consistently worsen (see Fig. \ref{fig:FricErrors}). 
 
 The modeling convention is to keep the Rayleigh frictional timescale fixed at a standard value (very often a 1 day timescale inherited from the Held-Suarez test) regardless of the rotation rate \citep{wang2018comparative,penn2018atmospheric} or vary it as a free parameter over arbitrarily large ranges \citep{pascale2013nonequilibrium,carone2016connecting}. But these results show that when the frictional timescale is equal to or larger than the rotational timescale, the model becomes unreliable.

 We also experimented with changing grid resolution and numerical dissipation order within in the model. A finer grid allows \added{for} the resolution of smaller scale flows, and a higher-order numerical diffusion is more scale selective in damping at the grid scale, both of which are desirable \citep{mendonca2016thor,skinner2020numerical}. For the high resolution run the spatial resolution was doubled and the model time step was proportionally \replaced{halved to roughly keep the Courant number the same, to keep}{halved to achieve an approximately equal Courant number, thereby keeping} the simulations amenable to intercomparison. The Courant number is calculated as:

 \begin{equation}
     C = \frac{U\Delta t}{d}
 \end{equation}

 where $U$ is the magnitude of horizontal velocity at a given grid point, $d$ is the grid spacing \citep[see Eqn.2 of ][]{mendonca2016thor} and $t$ is the model time step. Please see the Sec. \ref{sec:Courant} of the appendix for further Courant number distribution statistics.
 
\begin{figure*}
 \plotone{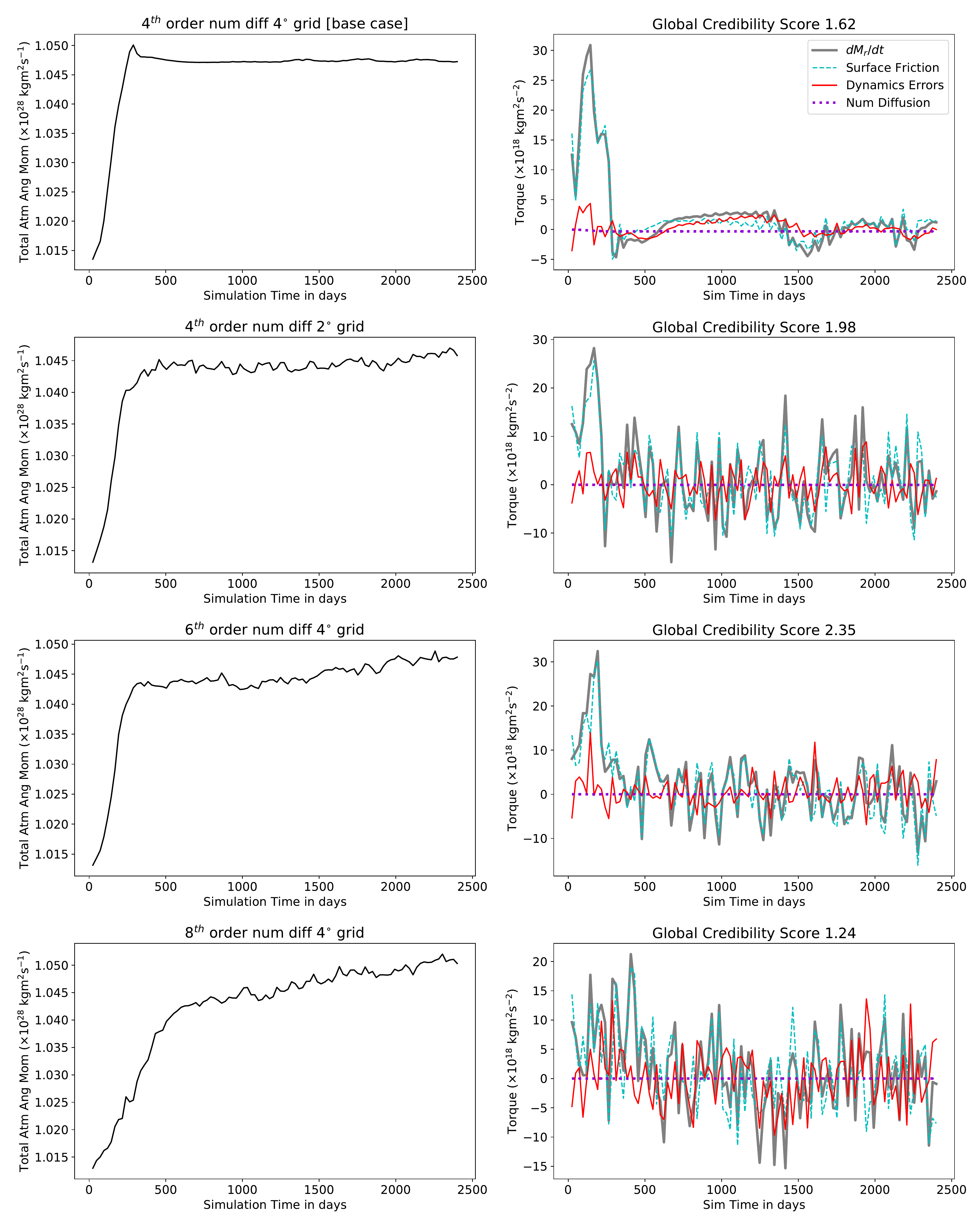}
 \caption{Same as Fig. \ref{fig:MainErrors} but with higher model grid resolution (second row) and higher numerical diffusion order (third and fourth rows) for a rotation rate of 1 day. \label{fig:DiffErrors}}
 \end{figure*}

 The numerical diffusion timescale, defined as \citep{mendonca2016thor}:
\begin{equation}
    \tau _d = \frac{\Delta t}{2^{(n+1)}D_{h}}
\end{equation}
was held constant (with a value of 3900 $s$) when changing the numerical diffusion order $n$ by adjusting the tuning parameter $D_{h}$ so that the simulations can be intercompared. As seen in Fig. \ref{fig:DiffErrors} increasing the resolution decreases error but increasing the order of the numerical diffusion scheme from 4 to 6 and further to 8 has mixed results. Furthermore, the eighth-order diffusion simulations take much longer to spin-up, though why this happens is as yet unknown.

The fact remains that all general circulation models must be tuned to be run stably and this tuning often takes the form of numerical diffusion, sponge layers and/or spectral filters. It is a very subjective choice in terms of how it is implemented, what its magnitude is and what the goal of the tuning is (e.g., to get a smooth kinetic energy spectrum or to have the smallest possible values of numerical diffusion that won't crash the model), see \citet{jablonowski2011pros} for a review of various tuning philosophies. But model tuning and model error are frequently under-discussed in exoplanet climate literature and often the tuning set-up for a test case is used unchanged for simulations across a wide range of physical parameters without checking for changes in model errors.

\section{Conclusions}
\label{sec:conc}

In this work, we took a model tuned to work at \replaced{one day rotation period}{a rotation period of one day} and examined its behavior at over four orders of magnitude in rotation rate, with a focus on angular momentum changes. We find that angular momentum changes are closely related to surface friction torques for the slow rotating planets, while model errors dominated the balance for the fast rotators. The sensitivity of model errors to numerical diffusion order, grid spacing and frictional timescale were also examined. The main takeaway is that model credibility can vary widely over a given parameter space. In short, for a given GCM if different settings show different variabilities, then this framework argues that you must trust the setting which yields the most physical angular momentum balance. The big challenge is to find tunings that allow for reasonably physical GCM runs across wide parameter spaces, but it is fairly straightforward to tune the model if you are interested in a single planet. Therefore, exoclimate parameter studies are much harder to get right than individual exoplanet modeling. Overall, it might be good to do an ensemble of different GCM runs to overcome individual GCM quirks when trying to estimate exoclimate variability, as has become standard practice in the Earth climate modeling community. \added{The THAI project is one such intercomparison effort focusing on terrestrial exoplanets \citep{fauchez2020trappist}.}

It is well understood within the modeling community that changing the model tuning parameters changes the resulting atmospheric flow. But a less recognized corollary is that each tuning is only appropriate for a given set of physical model parameters. By holding the tuning constant and varying the physical parameters, there is a risk of over or under-damping the numerical effects within the model, exactly as will happen by holding the physical parameters constant and changing the tuning. 

Since we have only tested one model within a relatively narrow parameter space, this work is intended to be indicative and not exhaustive.  \added{Atmospheric angular momentum conservation is considered an important metric in hot Jupiter simulations \citep{rauscher2010three,polichtchouk2014intercomparison,mayne2017results}, so our framework can be readily applied in those cases as well.} The magnitude of the model error is sensitive to the type of dynamical core (with finite-volume cores having much higher errors than spectral ones \citep{lauritzen2014held}\deleted{)} and both physical and model tuning parameters, as shown in this work. Other model error metrics based on conservation laws could be devised. We hope that exoclimate modeling efforts begin to include model error metrics when describing results, particularly for parameter space exploration type studies. 
\renewcommand{\thefigure}{A\arabic{figure}}

\setcounter{figure}{0}
\appendix
\section{Mean vs Median}
\begin{figure*}
 \plotone{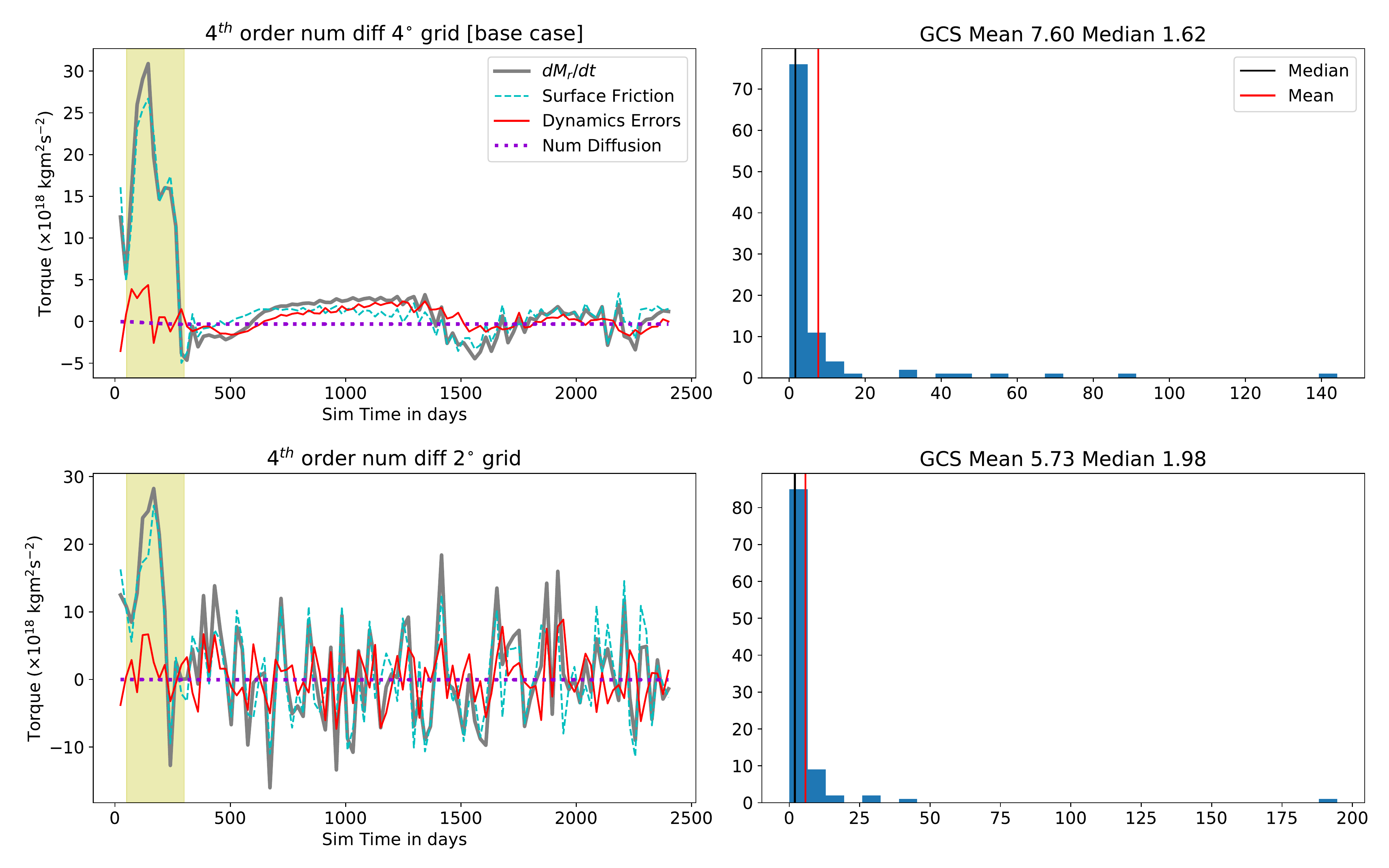}
 \caption{ The first two rows of \ref{fig:DiffErrors} are shown here with the histogram of the quantity $F/(D+\epsilon)$ calculated at each output timestep which is used to calculate the GCS through Eqn. \ref{eqn:gcs}.  The right panels show the mean and median of the distribution of these ratios, which were used to calculate two possible GCS values \label{fig:Histogram}}
 \end{figure*}

Here we show why the choice of median was necessary in the calculation of the credibility scores. In the right panels of Fig. \ref{fig:Histogram}, the distribution of the ratio of frictional torques to numerical torques has a long tail with a few extremely high values that occur during the model spin-up (indicated by the shaded region in the corresponding left panels). The mean is biased high because of these extreme values and not a good central measure of the distribution.  Therefore we use a median measure which is more resistant to such outliers.

\section{Courant number statistics}
\label{sec:Courant}
\begin{figure*}
 \plotone{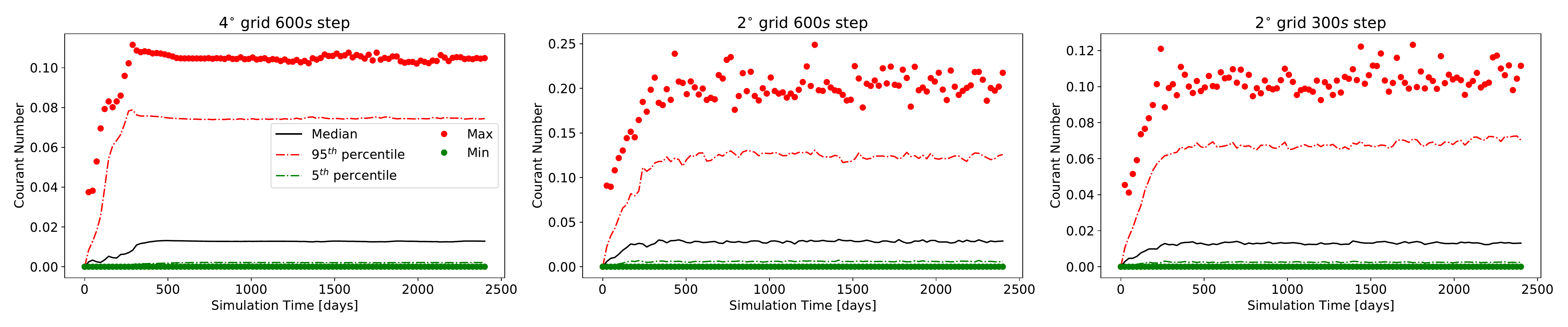}
 \caption{Distribution of Courant numbers for three runs with a 1 day rotation period and variable spatial and time resolution. The maximum and minimum values are also plotted at each timestep to make sure that there are no anomalous grid cells with very large values.  \label{fig:Courants}}
 \end{figure*}

Here we show further statistics for the Courant number with changes in grid resolution and time step. As shown in Fig. \ref{fig:Courants}, when the resolution in increased without adjusting the timestep (second panel), the Courant number spread is roughly twice as wide and is not a fair counterpart for intercomparison with the lower resolution run. When the timestep is halved (third panel) the Courant number range is similar to the lower resolution run, but the extreme values are more dispersive as higher resolution simulations allow for greater variability in velocities.

\acknowledgments

PK acknowledges funding support from the CSH Fellowship of the University of Bern.  RD and KH acknowledge funding support from the European Research Council via a Consolidator Grant awarded to KH (project EXOKLEIN; number 771620). KH acknowledges a honorary professorship from the University of Warwick.

%This work has been carried out within the framework of the NCCR PlanetS supported by the Swiss National Science Foundation.   KH: none of us is actually financed by the NCCR.

\newpage
\bibliography{errorrefs}{}
\bibliographystyle{aasjournal}



\end{document}